# Experimental demonstration of waveform-selective metasurface varying wireless communication characteristics at the same frequency band of 2.4 GHz

D. Ushikoshi, M. Tanikawa, K. Asano, K. Sanji, M. Ikeda, D. Anzai and H. Wakatsuchi

We report a waveform-selective metasurface that operates at 2.4 GHz band, i.e. one of ISM (Industrial, Scientific and Medical) bands. This structure gives us an additional degree of freedom to control electromagnetic waves and absorbs a particular waveform or pulse width of an incident wave at the same frequency band, while transmitting others. This unique capability is demonstrated using either ideal sinusoidal waves or Wi-Fi signals as a more practical waveform in wireless communications. Especially, this study shows how the waveform-selective metasurface interacts with realistic wireless communication signals from the viewpoint of communication characteristics, such as EVM (Error Vector Magnitude), BER (Bit Error Rate) and phase error. Thus, our study paves the way for extending the concept of waveform selectivity from a fundamental electromagnetic research field to a more realistic wireless communication field to, for instance, mitigate electromagnetic interference occurring at the same frequency band without significantly degrading communication characteristics.

*Introduction:* Electromagnetic interference is an important issue in recent years, as modern society is supported by various wireless communication devices (e.g. broadcasting antennas, smartphones, wireless LAN (local area network) routers and IoT (internet of things) devices, etc). These devices may be interfered by external electromagnetic fields, which leads to temporal malfunction or permanent damage [1]. Classically, this issue was addressed using, for instance, RF (Radio-Frequency) absorbers that converted the energy of an incoming electromagnetic wave to thermal energy [2], [3]. In this case, unnecessary scattering was effectively suppressed to protect sensitive electronic devices from electromagnetic noise. Particularly, use of artificially engineered periodic surfaces, or the so-called metasurfaces [4], enabled us to markedly reduce the entire design thickness with light weight, thereby readily solving electromagnetic interference occurring even in physically limited spaces [5], [6]. However, this issue becomes more complicated at ISM (Industrial, Scientific and Medical) bands, which are internationally standardised and used for many applications ranging from amateur radio, radars, Wi-Fi, Bluetooth, ZigBee and IoT to microwave ovens and semiconductor plasma etching. This indicates that electronic devices are more often exposed to electromagnetic noise in these bands than in others. Another important issue here is that these signals and noise may share the "*same*" frequency band. For instance, 2.4 GHz band is used for Wi-Fi (IEEE 802.11) [7], Bluetooth and microwave ovens. For this reason, there is strong demand for a new technology/technique that senses only a particular electromagnetic wave even at the same frequency band, while eliminating other unnecessary signals and noise.

In more recent years, a series of studies on circuit-based metasurfaces were reported to preferentially select a particular waveform, or pulse width, of an incident wave and dissipate the energy of other waves at the same frequency [8], [9]. Such a new capability was expected to give us an additional degree of freedom to control electromagnetic waves and design more ideal wireless communication environment with reduced interference. From the practical viewpoint, however, none of studies has yet to report how waveform-selective metasurfaces interact with realistic wireless communication signals including their communication characteristics.

For this reason, we design and evaluate a waveform-selective metasurface operating at 2.4 GHz band, i.e. one of ISM bands. This structure is demonstrated to vary its absorbing performance for ordinary sinusoidal waves as well as for more realistic Wi-Fi signals used for 2.4 GHz band. Moreover, we report the effect of the waveform-selective metasurface on the communication performance for investigating a possibility that waveform-selective metasurfaces can selectively transmit/eliminate the Wi-Fi signals according to the pulse width. Therefore, this study paves the way for extending the concept of waveform selectivity from a fundamental electromagnetic research field to a more realistic wireless communication field to, for instance, mitigate electromagnetic interference occurring at the same frequency band without significantly degrading communication characteristics.

*Theory, material and method:* Our waveform-selective metasurface was composed of periodic unit cells, each of which had a square conducting patch (with minor trimmings at edges) and ground plane as well as a dielectric substrate in between. Additionally, each unit cell contained several circuit components including a set of four diodes. These diodes were deployed between conductor edges (see Fig. 1*a*) to play the role of a diode bridge so that electric charges induced by an incoming wave were fully rectified to generate an infinite set of frequency components. However, most of the energy was at zero frequency as theoretically predicted from the Fourier series expansion of the rectified electric charges [10]. Besides, these charges were temporarily stored at a capacitor inside the diode bridge and then discharged to a parallel resistor, resulting in strong absorption for a short pulse. This absorbing performance, however, was reduced once the incident waveform changed to continuous wave (CW), since it fully charged up the capacitor. Therefore, this waveform-selective metasurface enables us to sense a particular type of waveform even at the same frequency. Note that the waveform-selective absorbing mechanism is obtained as long as the bandwidth of our structure is wider than that of the incoming wave.

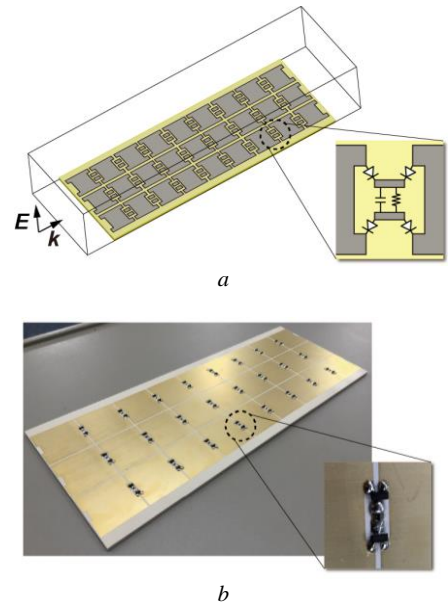

**Fig. 1** *Waveform-selective metasurface*
*a* The structure was deployed on the bottom surface of a standard rectangular waveguide (WR430)
*b* Measurement sample used

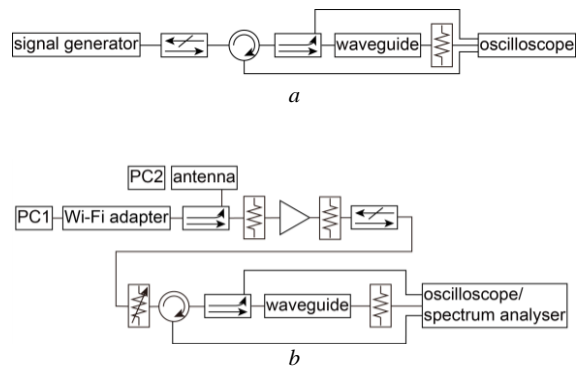

**Fig. 2** *Measurement setups*
*a* Measurement for scattering parameters
*b* Measurement for communication characteristics

In addition, while this study focuses on more effectively absorbing a short pulse than a CW by using the above capacitor-based waveform-selective metasurface, other types of structures can be alternatively used to more strongly absorb different waveforms such as CW and intermediate pulse than a short waveform [9].

Our measurement sample (Fig. 1*b*) had twenty seven square copper patches (each having the dimensions of 30 mm × 30 mm). These conductors formed a periodic array of 3 × 9 cells at 31 mm intervals on a dielectric substrate (Rogers3003, 3.04 mm thick) that was deployed on the bottom surface of a standard rectangular waveguide (WR430, Fig. 1*a*). Edges of copper patches were trimmed by 1.7 mm and 7.6 mm along the direction of the incident wave and the horizontal direction of the cross section of the waveguide, respectively, to deploy two small copper pads (2.4 mm by 2.0 mm each) that were used to connect circuit components. We used commercial schottky diodes provided by Broadcom (specifically, HSMS2863/2864). The capacitors and resistors inside diode bridges had 1 μF and 10 kΩ, respectively. Particularly, this capacitance value was large enough to sense the length of Wi-Fi signals used later.

Under these circumstances, the experimental sample was tested with either of the two measurement setups drawn in Fig. 2*a* and Fig. 2*b* to evaluate scattering parameters or communication characteristics. In the former setup, an incident wave was generated from a signal generator (Anritsu MG3692C). Scattering parameters were calculated using the energy of the waveforms observed in an oscilloscope (Keysight Technologies, DSOX6002A). In the latter setup, an incoming signal was generated from a commercial Wi-Fi adapter (Rosewill, RNX-N180UBEV2) that complied with IEEE 802.11b/g/n standards [7], [11]. As Wi-Fi signals differ in their waveforms every time, scattering parameters were obtained by averaging ten measurement data sets. To evaluate communication characteristics, the oscilloscope was exchanged with a spectrum analyser (Tektronix, RSA306B). In order to ensure the pulse width to be 50 μs long, we adjusted the payload size of each transmitted packet. The transmission timing including the pulse width was controlled on the Transport layer, so that, the signals used in this study totally abided by the IEEE 802.11-based Wi-Fi signals (i.e. no modification was made on PHY/MAC layers).

*Results and discussion:* Fig. 3 plots measurement results for the transmittance of the waveform-selective metasurface. This figure shows that with a low power level (e.g. with 0 dBm) the transmittance was independent of the incoming waveform and extremely small at 2.36 GHz. This limited transmittance is explained by the presence of a stop band, which was also seen in simulation (not shown here). By increasing the input power level to 10 dBm, however, the transmittance reduction seen in the short pulse shifted to a higher frequency, which is due to the absorbing mechanism for a short pulse. With a further increment in the power level (see 15 and 17 dBm in Fig. 3*a*), the transmittance reduction was mitigated, as the voltage across diodes approached their breakdown voltage, which allowed more electric charges to enter the diodes from their cathodes and thus lowered the waveform-selective absorbing mechanism. Note that these changes seen in the transmittance did not appear for a CW (Fig. 3*b*), because the waveform-selective metasurface behaved similarly to an ordinary stop-band metasurface.

These power dependences are more clearly plotted in Fig. 4, where the oscillating frequency was fixed at 2.4 GHz. According to this figure, the difference between the short-pulse and CW transmittances started increasing at 0 dBm and reached more than 10 dB at 15 dBm (see the closed symbols).

Additionally, this figure shows the power dependence for Wi-Fi signals (see the open symbols). Compared to simple pulsed sine waves, Wi-Fi signals had a wider bandwidth ranging from 2.402 to 2.422 GHz (refer to the grey areas of Fig. 3). Therefore, this difference led to overall improving the transmittances for both a short pulse and a CW. Moreover, the gap between the short-pulse transmittance and the CW transmittance reduced to a smaller value that also shifted to a lower power level, as the waveform-selective absorption/transmission had the dependences on both frequency and input power. However, these measurement results still ensure that our structure retains a waveform-selective performance larger than 6 dB for these Wi-Fi signals.

Let us discuss the communication performances controlled by the waveform-selective metasurface, such as the EVM (Error Vector Magnitude), BER (Bit Error Rate), and phase error characteristics demonstrated in Figs. 5 to 7, respectively. As can be seen from Fig. 5*a* and Fig. 5*b*, the difference in the EVM for the short pulse and CW with the waveform-selective metasurface became around 6 dB, which well agreed with the transmittance difference shown in Fig. 4 (see the open symbols). This means that the waveform-selective absorptance/transmission can successfully control the EVM in terms of communication performance. In addition, Fig. 6*a* to Fig. 6*d* show that the BER performance was also changed by the waveform-selective metasurface. In Fig. 6*a* and Fig. 6*b*, the theoretical BER was calculated under the assumption of AWGN (Additive White Gaussian Noise) channel. Note that the BER performance was degraded to over $10^{-2}$ in the case of the short pulse with the waveform-selective metasurface, which indicates that the short pulse was eliminated by the structure effectively. On the other hand, the phase errors for both short pulse and CW (Fig. 7*a* and Fig. 7*b*) varied in an extremely limited range smaller than ±1 degree. Therefore, the waveform-selective metasurface does not produce any large distortion for the Wi-Fi signals. Consequently, it can be concluded that the long Wi-Fi signals (CWs) can pass through the waveform-selective metasurface without any significant performance degradation, whereas the short Wi-Fi signals (pulse) can be effectively eliminated in terms of the communications performances.

This study focused on effectively reducing the transmittance and communication characteristics of short Wi-Fi signals. Note that those of long Wi-Fi signals can be alternatively lowered by using a different type of waveform-selective metasurface [9], [12]. In addition, although we used Wi-Fi signals as an example of realistic communication signals, the concept of waveform selectivity can be potentially applied to preferentially absorbing or transmitting other standardised wireless signals, as long as the bandwidths of waveform-selective metasurfaces are broader than those of the signals.

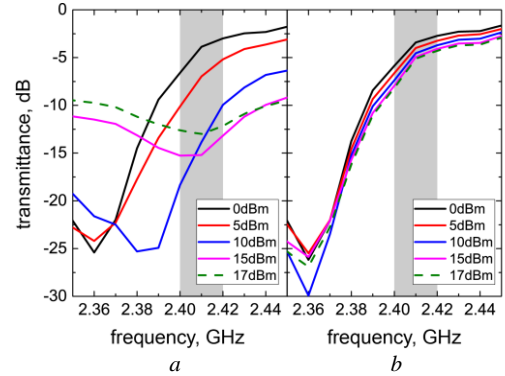

**Fig. 3** *Transmittance of the waveform-selective metasurface. The grey areas represent the bandwidth of the Wi-Fi signals used later in this study (see Fig. 4 to Fig. 7)*
*a* 50-μs pulse
*b* CW

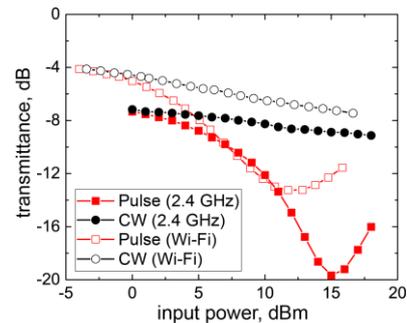

**Fig. 4** *Power dependence of transmittance of the waveform-selective metasurfaces for sine wave or Wi-Fi signal.*

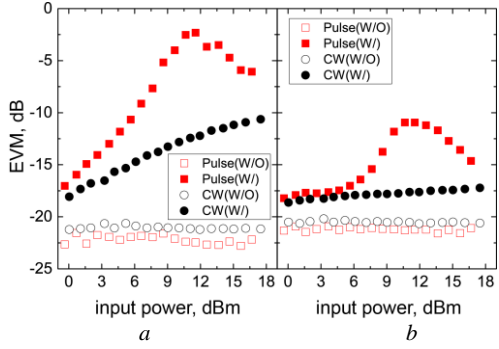

**Fig. 5** *EVM with or without the waveform-selective metasurface*
*a* Pilots
*b* Data

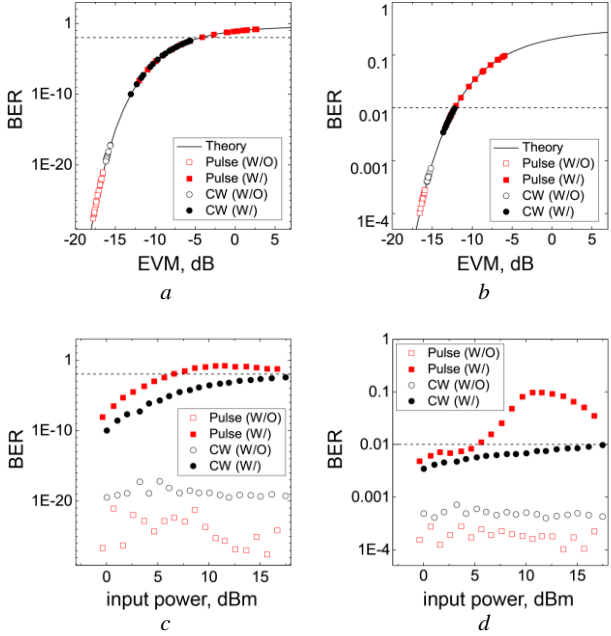

**Fig. 6** *BER with or without the waveform-selective metasurface. The dashed lines represent BER = 0.01*
*a* Pilots (as a function of EVM)
*b* Data (as a function of EVM)
*c* Pilots (as a function of input power)
*d* Data (as a function of input power)

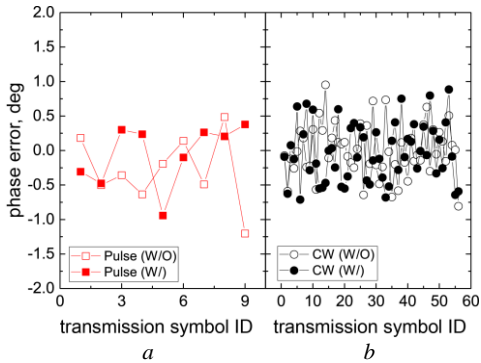

**Fig. 7** *Phase error with or without the waveform-selective metasurface*
*a* 50-μs pulse
*b* CW

*Conclusion:* We have reported performances of a waveform-selective metasurface working at 2.4 GHz band, which is known as one of ISM bands. The waveform-selective metasurface was experimentally tested to more strongly absorb a short pulse than a CW, which effectively varied transmittance at the same frequency of 2.4 GHz. Similar results were obtained when the incident source was changed to a commercial Wi-Fi adapter. In this case, the communication performances were able to be controlled according to the pulse width of Wi-Fi signals in the meaning of the EVM and BER characteristics. These results ensure that waveform-selective metasurfaces can be exploited not only for controlling simple scattering parameters but also for varying communication characteristics. Hence, our study paves the way for extending the concept of waveform selectivity from a fundamental electromagnetic research field to a more realistic wireless communication field, for instance, to mitigate electromagnetic interference occurring at the same frequency band without significantly degrading communication characteristics.

*Acknowledgments:* This work was supported by Denso.

D. Ushikoshi, M. Tanikawa, K. Asano, D. Anzai and H. Wakatsuchi (Department of Electrical and Mechanical Engineering, Graduate School of Engineering, Nagoya Institute of Technology, Aichi 466-8555, Japan)

E-mail: wakatsuchi.hiroki@nitech.ac.jp

K. Sanji and M. Ikeda (Research Department 23, Research & Development Department 2, SOKEN. INC., Aichi 470-0111, Japan)

H. Wakatsuchi: also with Precursory Research for Embryonic Science and Technology (PRESTO), Japan Science and Technology Agency (JST), Saitama 332-0012, Japan